\documentstyle[12pt]{article}
\newcommand{\be}{\begin{eqnarray}}
\newcommand{\ee}{\end{eqnarray}}
\begin{document}
\thispagestyle{empty}

\vspace*{4cm}

\begin{center}
{\large{\bf Zig Zag symmetry in AdS/CFT Duality\\ }}
\vspace{1.8cm}
{\large A.~I.~Karanikas and C.~N.~Ktorides}\\
\smallskip
{\it University of Athens, Physics Department\\
Nuclear \& Particle Physics Section\\
Panepistimiopolis, Ilisia GR 15771, Athens, Greece}\\
\vspace{1cm}

\end{center}
\vspace{0.4cm}

\begin{abstract}
The validity of the Bianchi identity, which is intimately
connected with the zig zag symmetry, is established, for piecewise
continuous contours, in the context of Polakov's gauge
field-string connection in the large 'tHooft coupling limit,
according to which the chromoelectric `string' propagates in five
dimensions with its ends attached  on a Wilson loop in four
dimensions. An explicit check in the wavy line approximation is
presented.
\end{abstract}
\newpage

\vspace*{.3cm}

{\bf 1. Introduction}

\vspace*{.2cm}

The employment of string theoretical methods to build inroads to
QCD, especially at non-perturbative level, is a problem which has
been posed by Polyakov [1] over two and a half decades ago. Since
then, string theory has made notable advancements in this regard,
both on applications to high energy processes [2] and in the
direction of expediting high order, perturbative computations;
see, e.g. [3] for a review presentation, wherein relevant aspects
to collider physics applications are also discussed; for recent
advances on this subject, see [4].

In an independent development and in the context of 'tHooft's [5]
large $N$, $\lambda\equiv g^2_{YM}N\gg 1$ limit, Polyakov [6]
proposed, in an attempt to capture the essential characteristics
of a string relevant to QCD and which accommodates the Liouville
mode, a setting according to which the string appropriate for
representing the chromo-electric flux lines of a pure Yang-Mills
theory must propagate in a five dimensional environment the metric
of which reads
\begin{equation}
ds^2=a(y)(dy^2+dx_\mu^2),\quad a(y)\sim y^{-2}\,\,(y\rightarrow
0),
\end{equation}
with the gauge theory `living' at the boundary, $y=0$, of this
space. The above description will contain additional dimensions,
if the 4-D theory has extra matter fields, as it happens in the
AdS/CFT case [7]. Conformal symmetry requirement fixes
\begin{equation}
ds^2=a(y)={R^2\over y^2},\quad R^2=\alpha'\sqrt{\lambda}.
\end{equation}

The Wilson loop functional [8]
 \begin{equation}
W[C]={1\over N}\langle Tr P\exp i\oint\limits_C A_\mu
dx_\mu\rangle_A
\end{equation}
plays a basic role in the gauge-string correspondence in
Polyakov's scheme wherein the open string propagating in a
5-dimensional background (1) has its two ends attached onto a loop
contour. The latter, as already mentioned, lives in 4-dimensions.

The working assumption for quantifying such a proposal is that, in
the large $\lambda$ limit, the Wilson loop functional is expected
to behave as
\begin{equation}
W[C]\propto e^{-\sqrt{\lambda}{\cal A}_{\min}(C)},
\end{equation}
where ${\cal A}_{\min}$ is the minimal area swept by the string
and is bounded by the contour $C$. This statement constitutes a
zeroth, WKB-type, approximation to the problem.

Now, the loop casting of QCD has a long history which is
intimately associated with theoretical efforts to probe its
nonperturbative content. It constitutes a well defined strategy of
formulating QCD and enjoys, in its discrete version, universal
acceptance as {\it the} methodology for investigating non
perturbative issues surrounding strong force dynamics.

A corresponding, direct continuum casting of QCD, based on the
Wilson functional, gives rise to the loop equation formalism which
has been extensively pursued by Makeenko, Migdal [9,10], as well
as by Polyakov in [1] and has provided a multitude of powerful
insights to the theory. Within the framework of this scheme, a
property of vital importance Wilson functionals must posses is
that of zig-zag, equivalently backtracking, invariance. The same
symmetry plays a fundamental role in Polyakov's choice of the
background (1) that accommodates the fluctuations of the random
surfaces bounded by the contour $C$. Such a requirement
characterizes, in general, the so-called Stokes type functionals
whose basic property is, precisely, that they do not change when a
small path passing back and forth is added to any smooth section
of the loop at any given point. In Mathematics, this property is
associated with what are known as Chen integrals. Quantitatively
speaking, the backtracking invariance in the loop formalism
assumes the form (see, e.g., [10]) \be
\epsilon^{\kappa\lambda\mu\nu}\partial^x_\lambda
\frac{\delta}{\delta \sigma_{\mu\nu}(x)}W[C]=0,
 \ee
 with $\delta \sigma_{\mu\nu}$ and $\partial_\lambda^x$ the surface and path
derivatives whose action will be specified later. From the point
of view of QCD the relevance of Stokes type functionals is traced
to the fact that they facilitate the proof of the non-abelian
Stokes theorem, hence their name.

In order to establish the validity of the non-Abelian Stokes
theorem in the loop formalism of QCD the key role is played by the
Bianchi identity, which assures the commutativity of
differentiations performed on a Wilson loop, in a surface
independent manner [10-12]. In fact, one easily verifies that
\begin{equation}
\epsilon^{\kappa\lambda\mu\nu}\partial_\lambda^x\frac{\delta}{\delta\sigma_{\mu\nu}}W={1\over
N}\epsilon^{\kappa\lambda\mu\nu}{\rm Tr} P\langle
\bigtriangledown_\lambda F_{\mu\nu}\exp i\oint\limits_C A_\mu
dx_\mu\rangle_A=0.
\end{equation}

Demonstrating the validity of the Bianchi identity, equivalently
zig zag invariance, within the framework of the field-string
connection according to the proposal in Ref. [6], is the central
objective of this work. More specifically, the stated objective of
this paper is to establish that
\[
\epsilon^{\kappa\lambda\mu\nu}\partial^x_\lambda\frac{\delta}
{\delta\sigma_{\mu\nu}}\exp(-\sqrt{\lambda} A_{\min})\approx 0,
\]
in the limit $\lambda\rightarrow \infty$.

Our exposition is organized as follows. In the next section we
introduce the area derivative operator appropriate for acting on
the Wilson loop functional. To begin with, on the field
theoretical side it is through this action that one establishes
the loop equations. On the string side, it will turn out that it
plays a key role in establishing the Bianchi identity. The
variational analysis for the verification of both the loop
equations and the Bianchi identity will be greatly facilitated by
employing a methodology, developed in Refs [13,14], which directly
addresses a situation involving a surface bounded by a closed
contour in four (D) dimensions that variationally protrudes into
five (D+1) dimensions. This approach will be reviewed in Section
2, where the all important quantity, to be designated as
$\vec{g}$-function, will emerge. This quantity, as it turns out,
contains all the dynamics in the advocated approach. The area
derivative operator will also be introduced in this section and
some realizations of general nature will be made regarding its
action on the Wilson loop functional. The next section (3) is
devoted to the study of the normal, with respect to the boundary
of the Wilson loop, variations of the $\vec{g}$ function. These
variations will play a pivotal role in our subsequent quantitative
considerations. In section 4 we apply the mathematical formalism
developed to this point to verify, on the string side, the loop
equation of Makeenko and Migdal [9]. At the same time we shall
derive a conditional, at this stage, result concerning the Bianchi
identity. The conditionality of the result will be attributed to
the fact that the vector basis adopted to describe the five
dimensional surface spanned by the string is too general to
control the precise manner by which it ``collapses'' onto the,
corresponding, four-dimensional Wilson loop configuration.
Accordingly, only a {\it condition} for the validity of the
Bianchi identity can be obtained. Full confirmation becomes
precise in Section 5 where a certain Wilson contour of sufficient
generality introduced in Ref. [13] and characterized as `wavy
line' configuration, is employed to rigorously demonstrate the
validity of the Bianchi identity. Some general, concluding
comments are presented in the final section.

\vspace*{.3cm}

 {\bf 2. String Action Functional and the area
derivative operator}

\vspace*{.2cm}

In this section we present the general form of the area derivative
operator which is to act on a Wilson loop configuration.  We begin
our discussion by presenting a condensed summary of the setting
promoted in Refs [13,14] which is nicely suited for conducting
analytical considerations pertaining to the proposal of Ref. [6].
The relevant string action functional according to this reference
is (Euclidean formalism employed throughout)
\begin{eqnarray}
S[\vec{x}(\xi),y(\xi)]&=&{1\over 2}\sqrt{\lambda}
\int_D d^2\xi G_{MN}(x(\xi))\partial_ax^M(\xi)\partial_a x^N(\xi)\nonumber\\
&=& {1\over 2}\sqrt{\lambda}\int_D \frac{d^2\xi}{y^2(\xi)}
[(\partial_a \vec{x}(\xi))^2+(\partial_a y(\xi))^2],
\end{eqnarray}
where $x^M=(y,\vec{x})=(y,x^\mu),\,
M,N=0,1,\cdot\cdot\cdot,D;\,\mu=1,\cdot\cdot\cdot,D$, with the
$y$-coordinate taking a zero value at the boundary and growing
toward infinity as one moves deeper into the interior of the
AdS$_5$ space\footnote{To connect, in a general sense, our present
work with the AdS/CFT conjecture [7], we shall, in a loose sense,
refer to the 5-dimensional space-time background of Polyakov's
scheme, wherein conformal invariance is implicitly assumed, as
AdS$_5$.}.

In Refs [13,14] a mathematical machinery was developed for the
purpose of studying loop dynamics in reference to the above action
functional. We shall adopt the strategy introduced in these works,
the immediate aim being to determine the action of the area
derivative operator [15]
 \be
\frac{\delta}{\delta\sigma_{\mu\nu}(x(\sigma))}=\lim\limits_{\eta\rightarrow
0} \int\limits_{-\eta}^\eta dh\,h\frac{\delta^2}{\delta
x_\mu\left(\sigma+{h\over 2}\right)\delta
x_\nu\left(\sigma-{h\over 2}\right)} \ee on a piecewise regular
Wilson loop contour.

The loop functional is to be minimized under the boundary
conditions $\vec{x}|_{\partial D}=\vec{c}(\alpha(\sigma))$ and
$y|_{\partial D}=0$, with the parametrization chosen so that
\begin{equation}
A_{\min}[\vec{c}(\sigma)]=\min\limits_{\{\alpha(\sigma)\}}\min\limits_{\{\vec{x},y\}}S[\vec{x}(\xi),y(\xi)],
\end{equation}
The functional $A_{\min}$ is invariant under reparametrizations of
the boundary, a property that can be easily deduced  from the
above minimization condition ($c'_\mu(s)={d\over ds}c_\mu(s)$):
\begin{equation}
c'_\mu(\sigma)\frac{\delta A_{\min}}{\delta c_\mu(\sigma)}=0.
\end{equation}

Following Refs [13,14], we adopt the static gauge $y(t,\sigma)=t$
and place the loop on the boundary of the AdS$_5$ space, i.e. set
$t= 0$. One, accordingly, writes
\begin{equation}
\vec{x}(t,\sigma)=\vec{c}(\sigma) +{1\over
2}\vec{f}(\sigma)t^2+{1\over 3}\vec{g}(\sigma)t^3+{1\over
4}\vec{h}(\sigma)t^4\cdot\cdot\cdot,
\end{equation}
where, for now, the curve $\vec{c}(\sigma)$ is assumed to be
everywhere differentiable. If there are cusps on the loop contour
({\it i.e.}, discontinuities in the first derivative) the above
expansion must be understood piecewise. Surface minimization leads
to the elimination of the linear term in the expansion and
determines its next coefficient:
\begin{equation}
\vec{f}=\frac{d}{d\sigma}\frac{\vec{c'}}{\vec{c'}^2}.
\end{equation}
The coefficient  $\vec{g}(\sigma)$ is, at this point, unspecified.
Imposition of the Virasoro constraints leads to
\begin{equation}
\vec{c'}\cdot\vec{g}=0.
\end{equation}
It turns out that the latter relation simply expresses the
reparametrization invariance of the minimal area (9) and, hence,
the quantity $\vec{g}(\sigma )$, to be referred to as
$\vec{g}$-function from hereon, remains undetermined. More
illuminating, for our purposes, is an interim result through which
(13) is derived and reads as follows
\begin{equation}
\frac{\delta
A_{\min}}{\delta\vec{c}(\sigma)}=-\sqrt{\vec{c'}^2}\vec{g}(\sigma).
\end{equation}
The above relation underlines the dynamical significance of the
$\vec{g}$-function: It provides a measure of the change of
$A_{\min}$ when the Wilson loop contour is altered as a result of
some interaction which reshapes its geometrical profile.

Consider, now, the action of the area derivative on the Wilson
loop functional:
\begin{equation}
\frac{\delta}{\delta
\sigma_{\mu\nu}(\sigma)}W[C]=\lim\limits_{\eta\rightarrow
0}\int\limits_{-\eta}^\eta dh\,h \left[-\sqrt{\lambda}
\frac{\delta^2 A_{\min}} {\delta c_\mu\left(\sigma+{h\over
2}\right){\delta c_\nu\left(\sigma-{h\over 2}\right)}}
+\lambda\frac{\delta A_{\min}} {\delta c_\mu\left(\sigma+{h\over
2}\right)} \,\frac{\delta A_{\min}} {\delta
c_\nu\left(\sigma-{h\over 2}\right)}\right]W[C].
\end{equation}
As it is known [16], the area derivative is a well defined
operation only for smooth contours, {\it i.e.} everywhere
differentiable ones. In such a case the last term in the above
equation gives zero contribution. If the loop under consideration
has cusps, as happens in the framework of non-trivial situations,
the operation must be understood piecewise; see Ref. [17] for such
a realization.

To further facilitate our considerations we follow Ref(s) [13,14]
by choosing the coordinate $\sigma$ on the minimal surface such
that
\[
\vec{c'}^2(\sigma)=1,\quad \dot{\vec{x}}(t,\sigma)\cdot
\vec{c}'(\sigma)=0.
\]
We also introduce an orthonormal basis, which adjusts itself along
the tangential ($\vec{t}$) and normal ($\vec{n}^a\,
,a=1,\cdot\cdot\cdot,D-1$) directions defined by the contour, as
follows
\begin{eqnarray}
&&\{\vec{t}, \vec{n}^a\},\,a=1,\cdot\cdot\cdot,D-1\nonumber\\
&&
\vec{t}=\frac{\vec{c}\,'}{\sqrt{\vec{c}\,^2}},\quad\vec{n}^a\cdot\vec{t}=0,\quad
\vec{n}^a \cdot\vec{n}^b=\delta^{ab}.
\end{eqnarray}
We now write
\begin{equation}
\frac{\delta}{\delta
c_\mu}=n_\mu^a\left(\vec{n}^a\cdot\frac{\delta}{\delta
\vec{c}}\right) +t_\mu\left(\vec{t}\cdot\frac{\delta}{\delta
\vec{c}}\right)\equiv n_\mu^a\frac{\delta}{\delta\vec{n}^a}
+t_\mu\frac{\delta}{\delta \vec{t}}
\end{equation}
and upon using relations (12) and (13), as well as setting
$s=\sigma+h/2$ and $s'=\sigma-h/2$,  we determine
\begin{equation}
\frac{\delta^2 A_{\min}} {\delta c_\mu(s)\delta
c_\nu(s')}=-\frac{\delta g^a(s)}{\delta\vec{n}^b(s')}
n_\mu^a(s)n_\nu^b(s')+R_{\mu\nu}(s,s')\delta'(s-s'),
\end{equation}
where
\begin{equation}
R_{\mu\nu}(s,s')=2\vec{g}(s)\cdot\vec{n}^a(s')t_\mu(s) n^a_\nu(s')
+ \vec{g}(s)\cdot\vec{t}(s')t_\mu(s) t_\nu(s') -
\vec{t}(s)\cdot\vec{n}^a(s')g_\mu(s)n_\nu^a(s').
\end{equation}
From the defining expression, cf. Eq (8), one immediately realizes
that only terms $\sim \delta '(s-s')$ in an antisymmetric
combination $R_{[\mu\nu]}$ will give non-zero contributions to the
area derivative. It, thus, becomes obvious that the last term in
Eq (18) produces the result
\begin{equation}
R_{[\mu\nu]}(\sigma,\sigma) = t_{[\mu}(\sigma) g_{\nu]}(\sigma).
\end{equation}

Turning our attention to the first term on the rhs of (18) we note
that non-vanishing contributions should have the form
\begin{equation}
(r^aq^b-r^bq^a)n_\mu^an_\nu^b\delta'(s-s'),
\end{equation}
where $r^a=\vec{n}^a\cdot\vec{r}$ and $q^a=\vec{n}^a\cdot\vec{q}$.
These functions must be determined from the coefficients of the
expansion (11); otherwise the above contribution would be contour
independent, having no impact on a calculation associated with
non-trivial dynamics. In conclusion, a simple qualitative
analysis, based on the scale invariance of $A_{\min}$, indicates
that a contribution of the type (21) does not exist. This
qualitative observation can be further substantiated through a
straightforward argument based on dimensional grounds. Indeed,
from Eq. (11) it can be observed that under a change of scale of
the form $\vec{c}\rightarrow
\lambda\vec{c},\,(t,\sigma)\rightarrow (\lambda t,\lambda \sigma)$
one has
\[
\vec{c}'\rightarrow \vec{c}',\quad
\vec{f}\rightarrow{1\over\lambda} \vec{f},\quad \vec{g}\rightarrow
{1\over \lambda^2}\vec{g},\cdot\cdot\cdot.
\]
On the other hand, now, the area derivative, being of second
order, should scale $\sim {1\over \lambda^2}$. In turn, this means
that one of the quantities $\vec{r}$ or $\vec{q}$, which must
arise through transverse variations of $\vec{g}$, should be
aligned with the tangential vector $\vec{t}$ which , by
definition, has zero transverse components. Thus, the only
antisymmetric combination with the right scaling behavior must be
either of the form $r^af'^b-r^bf'^a$, or $r^ag^b-r^bg^a$, where
$r^a\sim n^a_ic'_i$, with $i=2,\cdot\cdot\cdot$. But such
expressions must be excluded since they pick out a certain
direction in the four dimensional space, whereas the area
derivative must be a second rank tensor.

Referring to the formula for the area derivative, one immediately
surmises that the first term on the rhs  of Eq. (18) gives null
contribution since the antisymmetric term is proportional to
$\delta(s-s')$, and {\it not} $\delta'(s-s')$. We have, therefore,
determined that
\begin{equation}
\lim\limits_{\eta\rightarrow 0}\int\limits_{-\eta}^\eta dh\,h
\frac{\delta^2 A_{\min}} {\delta c_\mu\left(\sigma+{h\over
2}\right){\delta c_\nu\left(\sigma-{h\over 2}\right)}}
=t_{[\mu}(\sigma)g_{\nu ]}(\sigma).
\end{equation}

In order to check the validity of the Bianchi identity we need a
quantitative expression of the normal, with respect to the
four(D)-dimensional surface of the Wilson loop, variations of the
$\vec{g}$-function. As it will turn out, the antisymmetric part of
the variations will play a determining role in the derivation of
the the Bianchi identity. A quantitative study of these normal
deviations will be conducted in the next section and the relevant
results will further justify the line of arguments promoted in
this section.

\vspace*{.3cm}

{\bf 3. The Normal Variation of the $\vec{g}$-function}

\vspace*{.2cm}

We start the considerations in this section by remarking that path
derivative entering the Bianchi identity can be defined by [10]
\begin{equation}
\partial_\mu^{c(s)}=\lim\limits_{\epsilon\rightarrow
0}\int\limits_{s-\epsilon}^{s+\epsilon}ds' \frac{\delta}{\delta
c_\mu(s')}.
\end{equation}
Accordingly, as it becomes obvious from Eq. (22) of the previous
section, one needs an explicit expression of the normal variations
of the $\vec{g}$-function. In fact, their antisymmetric part, as
it will turn out, will play the deciding role concerning the
eventual derivation of the Bianchi identity is concerned, as will
be explicitly established in the sections to follow.

Let us introduce at every point of the surface bounded by the
loop, a basis $\{n^a_M(t,s)\}$ of $D-1$ orthonormal vectors which
satisfy the conditions \be n_M^a (t,s)\dot x_M (t,s) = n_M^a
(t,s)x'_M (t,s) = 0, \ee where $G_{MN}n_M^a n_N^b=\delta^{ab}$ and
$n_\mu^a(0,s)=n_\mu^a(s)$.

Under the normal variation \be x_M (t,s) \to x_M (t,s) + \psi _M
(t,s),\quad \psi _M (t,s) = \phi ^a (t,s)n_M^a (t,s) \ee the
change of the minimal surface to second order in $\phi^a$ reads
\be S^{(2)}  = \int {d^2 } \xi \left[ {\sqrt g (g^{\alpha \beta }
\partial _\alpha  \psi ^a
\partial _\beta  \psi ^a  + 2g^{\alpha \beta } \omega _\alpha ^{[ab]} \partial _\beta
\psi ^a \psi ^b  + 2\psi ^a \psi ^a ) + O(t^2 \psi ^2 )} \right]
\ee where we have written $\psi^a\equiv t\phi^a$ and have
introduced $g_{\alpha\beta}=G_{MN}
\partial_\alpha x_M\partial_\beta x_N$, while the, antisymmetric, quantities $\omega _\alpha ^{[ab]}$ are spin connection
coefficients and are given by \be \omega _\alpha ^{[ab]}  =
\partial _\alpha  n_M^a  \cdot n_M^a \ee
Details of the analysis can be found in [14]. Here, all we need is
the third order term in an expansion of $\psi_M$ in powers of $t$.
Notice that by taking into account that $\phi$ is regular as $t\to
0$, we have omitted terms $\sim t^4$ in (26) which do not
contribute to the normal variation of the $\vec{g}$-function.

Using the expansion (11) one easily determines that \be g_{\alpha
\beta }  = {1 \over {t^2 }}\left(
\begin {array}{ll}
  1 + \vec f^2 t^2+ 2\vec f \cdot \vec gt^3
  \quad\quad\quad\quad\quad\quad{1\over 2}\vec f \cdot \vec f't^3\\
  \quad\quad{1\over 2}\vec{f} \cdot \vec {f}'t^3 {\rm  }
  \quad\quad\quad\quad 1 - {{\rm 1} \over {\rm 2}}\vec f^2 t^2  -
{2 \over 3}\vec f \cdot \vec{g}t^3   + O(t^2 )\\
\end{array}\right)
\ee and \be \sqrt g  = {1 \over {t^2 }}(1 + {2 \over 3}\vec f
\cdot \vec gt^3 ) + O(t^2 ). \ee

Now, the area derivative receives contributions from antisymmetric
terms. We, therefore, have to find the behavior of the spin
connection as $t\to 0$. This cannot be done in a unique way if
$D>2$. What one can do is to expand the basis vectors $n_M^a(t,s)$
as a power series in $t$: \be
&&n_0^a (t,s)= tk_0^a (s)+{1 \over 2}t^2 l_0^a (s) + {1 \over 3}t^3 m_0^a (s) +\cdot\cdot\cdot\nonumber\\
&&\vec n_{}^a (t,s) = t\vec k_{}^a (s)+{1 \over 2}t^2 \vec l_{}^a
(s) + {1 \over 3}t^3 \vec m_{}^a (s) +\cdot\cdot\cdot \ee

Combining these relations with (24) and using the expansion (11)
we can determine that \be k_0^a  = f^a, \,\,l_0^a =-2(\vec k^a
\cdot \vec f + g^a ),\,\,m_0^a= - 3({1 \over 2}\vec l^a  \cdot
\vec f + \vec k^a \cdot \vec g + h^a ) \ee and \be \vec k^a \cdot
\vec c'=0,\,\,\vec l^a  \cdot \vec c'+ f^{'a} =0,\,\, \vec m^a
\cdot \vec c' + g^{'a}  + {3 \over 2}\vec k^a \cdot \vec f = 0.
\ee

From the orthonormality condition we find that \be
 &&\vec k^a  \cdot \vec n^b (s) + \vec k^b  \cdot \vec n^a (s) = 0,\,\,
2k_M^a  \cdot k_M^b  + \vec l^a  \cdot \vec n^b (s) + \vec l^b  \cdot \vec n^a (s) = 0\nonumber\\
&&\quad\quad\quad\quad {3 \over 2}l_M^a  \cdot l_M^b  + \vec m^a
\cdot \vec n^b (s) + \vec m^b  \cdot \vec n^a (s) = 0. \ee

With the above in place we return to our central objective and, to
start with, assume that \be \vec k^a  \cdot \vec c' = 0{\rm   }
\to {\rm   }\vec k^a  = \vec 0, \ee which means that \be
\begin{array}{cc}
\vec l^a  \cdot \vec c' =  - f'^a\\
\vec l^a  \cdot \vec n^b (s) + \vec l^b  \cdot \vec n^a (s) =  - 2k_0^a k_0^b  =  - 2f^a f^b.\\
\end{array}
\ee

From these relations we conclude that \be &&{\rm   }\vec l^a  =  -
f'{^a} \vec c' - f^a \vec f + \Lambda ^{ab} \vec n^b
(s)\nonumber\\ &&\vec{m}^a=-g'{^a}\vec{c}'-{3\over
2}(g^a\vec{f}+f^a\vec{g}) +M^{ab}\vec{n}^b(s), \ee with
$\Lambda^{ab}$, $M^{ab}$ antisymmetric, but otherwise arbitrary.

The first one, $\Lambda^{ab}$, enters the second order term in the
expansion (30) and consequently contributes to the normal
variation of the $\vec{g}$-function and through it to the area
derivative. The observation here is that this function cannot be
exclusively determined  from the functions
$\vec{c}',\,\vec{f},\,\vec{g},\cdot\cdot\cdot$ which, in turn,
determine $A_{\min}$. This can be deduced, through scaling
properties as follows: Under a change of scale
$\vec{c}\to\lambda\vec{c}, (t,s)\to\lambda(t,s)$, it must behave
as $\Lambda\to{1\over \lambda^2}\Lambda$, as can be seen from Eq.
(30). Taking, now, into account that
$\vec{c}'\to\vec{c}',\,\vec{f}\to{1\over \lambda}\vec{f},
\,\vec{g}\to{1\over \lambda^2}\vec{g},\cdot\cdot\cdot$ and that
$\vec{n}^a(s)\cdot\vec{c'}=0\to c'{^a}=0$, it becomes obvious that
it is impossible to find an antisymmetric combination of the
coefficient functions with the correct scaling behavior. The same
reasoning, in fact, justifies, {\it a posteriori}, Eq. (34). The
remaining possibilities are $\Lambda^{ab}=r^ag^b-r^bg^a$ or
$\Lambda^{ab}=r^af'^b-r^bf'^a$, with
$r^a=n^a_ic'_i,\,\,i=2,\cdot\cdot\cdot,D$. But, they are excluded
because the produced $l_\mu^a$ are not four dimensional vectors.
The second quantity, $M^{ab}$, must scale as $M^{ab}\rightarrow
{1\over \lambda^3}M^{ab}$ and consequently $M^{ab}\sim
g^af^b-g^bf^a$. Through this analysis the basis vectors are
determined as follows:
\begin{eqnarray}
n_0^a(t,s)&=&-tf^a-t^2g^a-t^3(h^a-f^a\vec{f}')+{\cal O}(t^4)
\nonumber\\ \vec{n}^a(t,s)&=& \vec{n}^a(s) -{1\over
2}t^3(g^a\vec{f}+f^a\vec{g})+{2\over
3}t^3(g'^a\vec{f}+f^a\vec{g}+{2\over 3}g'^a\vec{c'})
\nonumber\\&=&+{1\over 3}t^3\vec{n}^a M^{ab}+{\cal O}(t^4)
\end{eqnarray}

For the behavior of the spin connection we also need the
derivative $\vec{n}'{^a}(s)$. What we know about it comes from the
orthonormality condition
\begin{equation}
\vec n^a (s) \cdot \vec c'=0\to -\vec{n}'^a(s)\cdot \vec{c}=
-\vec{n}^a(s)\cdot\vec{c}''(s)=-\vec{c}''^a(s).
\end{equation}
Adopting the same arguments as before we conclude from the
preceding relation that
\begin{equation}
\vec {n}^{'}a (s) = - (\vec{n}^a (s) \cdot \vec {c}'')\vec {c}' =
- {c}{''}^a \vec {c}\,'
\end{equation}

In conclusion, through the above analysis we have determined that
\begin{equation}
\omega _t^{[ab]}={1 \over 2}t^2 \kappa_o(g^a f^b- g^b f^a
)\equiv{1 \over 2}t^2 r^{ab},\quad\omega _s^{[ab]} = {\cal O}(t^3
),
\end{equation}
with the constant $\kappa_o$ remaining undetermined at the present
level of the calculation.

Knowing the behavior of all the terms we now return to (26) and
demand the perturbed surface also to be minimal. This leads to the
equation
\begin{equation}
 \partial _\beta  (\sqrt g g^{\alpha \beta } \partial _\alpha  \psi ^a )
- 2\sqrt g \psi ^a  + 2\sqrt g g^{\alpha \beta } \omega _\alpha
^{[ab]} \partial _\beta  \psi ^b  = {\cal O}(t^2 \psi )
\end{equation}
To solve this equation we start from its asymptotic form as $t\to
0$, treating the other terms as small perturbations. At this point
it becomes very convenient to introduce, following Refs [13,14],
the Fourier transform
\begin{equation}
\phi ^a (t,s) = \phi ^a (t,s' + h) = \int\limits_{ - \infty
}^\infty  {{{dp} \over {2\pi }}} e^{iph} \tilde \phi ^a (t,p),
\end{equation}
with $s=\sigma+{h\over 2},\,s'=\sigma'-{h\over 2}$, the point at
which the area derivative is applied. The relevant observation
here is that one is interested in large values for the variable
$p\sim{1\over h}$, since the variable $h$ is integrated in the
vicinity of zero, {\it c.f.} Eq. (8).

On the other hand, one can be convinced, by appealing to (41),
that the values of $t$ which are involved in our analysis are
$t\sim {1\over |p|}\sim h$. With these estimations (40) can be
rewritten by retaining only those terms that are relevant to the
normal variation of the $\vec{g}$-function. To accomplish this
task the coefficient functions must be expanded around the point
$s'$. The general form of such an expansion can be read from
\begin{equation}
\begin{array}{cc}
F(s) = F(s') + (s - s')F'(s') + ... = F(s') + hF'(s') + ...,\\
h\phi ^a (t,s) = \int\limits_{ - \infty }^\infty  {{{dp} \over
{2\pi }}e^{iph} } h\tilde \phi ^a (t,p)
= \int\limits_{ - \infty }^\infty  {{{dp} \over {2\pi }}e^{iph} } i\partial _p \tilde \phi ^a (t,p).\\
\end{array}
\end{equation}
Given the above, Eq. (40) reads, in Fourier space,
\begin{equation}
\hat L_4^{ab} (t,p)\tilde \phi ^b (t,p)=\hat L_2^{ab}(t,p)\tilde
\phi ^b (t,p) + \hat L_1^{ab} (t,p)\tilde \phi ^b (t,p) + ...,
\end{equation}
where we have written
\begin{equation}
\begin{array}{cc}
 \hat L_4^{ab}  \equiv ({1 \over {t^2 }}\partial _t^2  - {2 \over t}\partial _t  - {{p^2 } \over {t^2 }})
\delta ^{ab},\quad \hat L_2^{ab}  \equiv \vec f^2 (\partial _t^2  + p^2 )\delta ^{ab}.\\
\hat L_1^{ab}  \equiv \left\{ {\left[ {2\vec f \cdot \vec
f'i\partial _p  + {4 \over 3}t(\vec f \cdot \vec g)}
\right](\partial _t^2  + p^2 ) + {4 \over 3}\vec f \cdot \vec
g\partial _t - {3 \over 2}\vec f \cdot \vec f'ip + t\vec f \cdot
\vec f'ip\partial _t } \right\}\delta ^{ab}
+ r^{ab} ({1 \over t} - \partial _t )\\
\end{array}
\end{equation}
The subscripts labelling the operators in the above relation serve
to signify their asymptotic behavior as $|p|\to\infty$:
\begin{equation}
\hat L_4^{ab} \tilde \phi ^b  \sim O(p^4 ),\quad\hat L_2^{ab}
\tilde \phi ^b \sim O(p^2 ),\quad\hat L_1^{ab} \tilde \phi ^b \sim
O(p).
\end{equation}
The neglected terms in (44) are of order ${\cal O}(p^0)$ so than
their contribution will be four times weaker that the strongest
one and thus irrelevant as far as we are interested in the normal
variation of the $\vec{g}$-function.

The solution of (44) can be written as
 \begin{equation}
\tilde \phi ^a (t,p) = \tilde \phi _{(0)}^a (t,p) +
\int\limits_0^\infty {dt'} G_p (t,t')\left[ {\hat L_2^{ab} (t',p)
+ \hat L_1^{ab} (t',p)} \right]\tilde \phi ^a (t',p).
\end{equation}
Here $\tilde{\phi}^a_{(0)}$ is the solution of the homogeneous
equation
\begin{equation}
\begin {array}{cc}
\hat L_4^{ab} (t,p)\tilde \phi ^b (t,p) = 0\\
\tilde \phi _{(0)}^a (t,p) = (1 + t\left| p \right|)e^{ - t\left| p \right|} \tilde \phi _{(0)}^a (p).\\
\end{array}
\end{equation}

The Green's function
\begin{equation}
\hat L_4^{ab} (t,p)G_p (t,t') = \delta (t - t')
\end{equation}
can be easily found:
\begin{equation}
G_p (t,t') = {1 \over {2\left| p \right|^3 }}\phi _ -  (t'\left| p
\right|)[\phi _ +  (t'\left| p \right|) - \phi _ -  (t'\left| p
\right|)]\theta (t - t') + (t \leftrightarrow t'),
\end{equation}
with
\begin{equation}
\phi _ -  (x)= (1 + x)e^{- x},\quad\phi _ + (x) = (1 - x)e^x.
\end{equation}
The solution of the integral equation (47) can be approached
through an iterative procedure:
\begin{equation}
\tilde \phi ^a (t,p) = \tilde \phi _{(0)}^a (t,p) +
\int\limits_0^\infty  {dt'} G_p (t,t')\left[ {\hat L_2^{ab} (t',p)
+ \hat L_1^{ab} (t',p)} \right]\tilde \phi _{(0)}^a (t',p) + {\rm
negligible\,\, terms}
\end{equation}

Expanding, now the result in a $t$-power series one can see that
the neglected terms in the above equation are of order ${\cal
O}(t^4)$ and thus irrelevant for our purposes. The symmetric part
of the solution (52) is easily determined to be
\begin{equation}
\left[ {1 - {1 \over 2}\left| p \right|^2 t^2  - {1 \over 3}t^3
(\vec f^2 \left| p \right| + i\vec f \cdot \vec f'signp + \vec f
\cdot \vec g)} \right]\tilde \phi _{(0)}^a (p) + O(t^4 ),
\end{equation}
while the contribution to the antisymmetric part is
\begin{equation}
\int\limits_0^\infty  {dt'} G_p (t,t')({1 \over {t'}} - \partial
_{t'} )e^{ - \left| p \right|t'} (1 + \left| p \right|t')r^{ab}
\tilde \phi ^a  =  - {1 \over 3}t^3 [\Gamma (0,2\left| p \right|t)
+ {{25} \over {12}}]r^{ab} \tilde \phi ^a  + O(t^4 ).
\end{equation}
The next step is to integrate the `annoying' incomplete gamma
function:
\begin{equation}
\int\limits_{ - \infty }^\infty  {{{dp} \over {2\pi }}} e^{iph}
\Gamma (0,2t\left| p \right|) = 2{\mathop{\rm Re}\nolimits}
\mathop {\lim }\limits_{\varepsilon  \to 0} \int\limits_0^\infty
{dp} e^{iph} \Gamma (\varepsilon ,2t\left| p \right|) =
2{\mathop{\rm Re}\nolimits} \mathop {\lim }\limits_{\varepsilon
\to 0} {t \over {2ih}}\Gamma (\varepsilon ) [1 - {1 \over {(1 +
{{ih} \over {2t}})^\varepsilon  }}] = {1 \over t}+{\cal O}(h)
\end{equation}
and thus the ${\cal O}(t^3)$ antisymmetric contribution to the
solution can be taken to be just
\begin{equation}
- {1 \over 3}t^3 {{25} \over {12}}r^{ab}=-{1\over 3}t^3\kappa
(g^af^b-g^bf^a).
\end{equation}
To obtain the final result one must take into account that normal
variations do not preserve the static gauge and, therefore, a
redefinition of the $t$ variable is needed. Repeating the relevant
calculation of Ref [13] we arrive at the following key result for
the normal variations of the components of the $\vec{g}$-function
\begin{eqnarray}
&&\frac{\delta g^a(s)}{\delta \vec{n}^b(s')}=\int {dp\over
2\pi}\left[\mid p\mid^3 -\mid p\mid
\left(\vec{f}^2\delta^{ab}-3f^af^b\right)\right] e^{iph}\nonumber\\
&& -\left[\vec{f}\cdot\vec{g}\delta^{ab}-{3\over
2}\left(f^ag^b+f^bg^a\right)+\kappa\left(f^ag^b-f^bg^a\right)\right]\delta(h)+{\cal
O}(h).
\end{eqnarray}

It should be stressed, at this point, that the arbitrariness of
the number $\kappa$ appearing in in Eqs (56) and (57) is related
to the the arbitrary number $\kappa_0$ that appears in Eq (40) by
$\kappa={25\over 12}\kappa_0$. The origin of this arbitrariness is
the fact that one cannot define uniquely an orthonormal basis on
the 5-dimensional surface.\footnote{The freedom of choosing of
such a basis was ignored in a previous work, namely
hep-th/0608030, where $\kappa_0$ was arbitrarily set to 1.}

\vspace*{.3cm}

{\bf 4. Loop equation and Bianchi identity}

Beginning this section we perform a first check of (22) by using
it to verify the Makeenko-Migdal (MM) equation [9], see also
extensive review expositions in Refs. [10], for  {\it
differentiable}, non self-intersecting Wilson loops which are
traversed only once, namely
\begin{equation}
\tilde{\Delta}W[C]\approx 0,
\end{equation}
where the symbol $\approx$ means that the finite part on the rhs
is zero and the MM loop operator is defined in [10] as
\begin{equation}
\tilde{\Delta}=\oint\limits_C dc_\nu\partial_\mu^c
\frac{\delta}{\delta \sigma_{\mu\nu}(c)}=
\lim\limits_{\eta\rightarrow 0}\lim\limits_{\eta'\rightarrow
0}\int ds\,c'_\nu(s) \int\limits_{s -\eta}^{s +\eta}
ds'\frac{\delta}{\delta c_\mu(s')}\int\limits_{-\eta'}^{\eta'}
dh\,h\frac{\delta^2}{\delta c_\mu(\ s +h)\delta c_\nu(s)}.
\end{equation}
It can, now, be easily determined from Eq. (22) that
\begin{equation}
\tilde{\Delta}A_{\min}=2\lim\limits_{\eta\rightarrow 0}\int
ds\,c'_\nu (s)\int\limits_{\ s-\eta}^{s+\eta}
ds'\frac{\delta}{\delta
c_\mu(s')}[t_{\nu}(s)g_\mu(s)]=2\lim\limits_{\eta\rightarrow 0}
\int ds\int\limits_{s-\eta}^{s+\eta}ds' \frac{\delta
g_\mu(s)}{\delta c_\mu(s')}.
\end{equation}
From Eq (18) we obtain
 \be
\frac{\delta g_\mu(s)}{\delta c_\nu(s')}= \frac{\delta
g^a(s)}{\delta
\vec{n}^b(s')}n_\mu^a(s)n_\nu^b(s')-R_{\mu\nu}(s,s')\delta'(s-s')-g_\mu(s)t_\nu(s)\delta'(s-s')
\ee One can easily check that $R'_{\mu\mu}(s,s)=0$ and
consequently \be
\tilde{\Delta}A_{\min}=2\lim\limits_{\eta\rightarrow 0}\int
\frac{\delta g^a(s)}{\delta
\vec{n}^b(s')}\vec{n}^a(s)\cdot\vec{n}^b(s')
\ee

From Eq (57) we see that
\begin{eqnarray}
&&\frac{\delta g^a(s)}{\delta
\vec{n}^b(s')}\vec{n}^a(s)\cdot \vec{n}^b(s')=-(D-4)\vec{f}\cdot\vec{g}\delta(s-s')+\nonumber\\
&&+\left[\frac{3!}{\pi}\frac{\delta^{ab}}{(\ s-s')^4}+{1\over
\pi}\frac{1}{(s-s')^2} (\vec{f}^2\delta^{ab}-3f^af^b)\right]
\vec{n}^a(s)\cdot\vec{n}^b(s')+{\cal O}(s-s')
\end{eqnarray}
and so, in a four dimensional space,

\be \tilde{\Delta}A_{\min}\equiv 0. \ee

It is obvious from the derivation of the above result that we
don't need to know the antisymmetric part of the normal deviations
of the $\vec{g}$-function for the verification of the MM loop
equation. This means that the fact that the numerical value of
$\kappa$ is unknown is of no importance, as far as the
verification of the loop equation is concerned. By juxtaposition,
for the verification of the Bianchi the antisymmetric part of Eq.
(57) plays a crucial role as we shall now witness.

To this end let us refer to Eq. (18) through which we find that
\begin{eqnarray}
&&t_\mu(s)\frac{\delta g_\nu(s)}{\delta
c_\lambda(s')}-(\mu\leftrightarrow \nu)=\frac{\delta
g^a(s)}{\delta\vec{n}^b(s')}
n_\lambda^b(s')t_{[\mu}(s)n^a_{\nu]}(s)\nonumber\\
\quad\quad&&+\delta'(s-s')\vec{t}(s)\cdot
\vec{n}^a(s')n_\lambda^a(s')t_{[\mu}(s)g_{\nu]}(s),
\end{eqnarray}
which finally gives
\begin{equation}
\epsilon^{\kappa\lambda\mu\nu}\partial^{c(s)}_\lambda \frac{\delta
A_{\min}}{\delta\sigma_{\mu\nu}(c(s))}=\epsilon^{\kappa\lambda\mu\nu}
\lim\limits_{\eta\rightarrow 0}\int\limits_{s-\eta}^{s+\eta}
ds'\frac{\delta
g^a(s)}{\delta\vec{n}^b(s')}n_\lambda^b(s')t_{[\mu}(s)n^a_{\nu]}(s)+\epsilon^{\kappa\lambda\mu\nu}
\vec{t}(s)\cdot\vec{n}'^a(s) n_\lambda^a t_{\mu}(s) g_{\nu}(s).
\end{equation}
One observes that, in the first term of the above equation only
the antisymmetric part of the normal variation of
$\vec{g}$-function survives. As far as the second term is
concerned,we can use the arguments presented in the previous
section to write $\vec{n}'^a=-(\vec{n}^a\cdot \vec{f})\vec{t}$.
The result expressed by Eq. (57) leads us now to conclude that
\begin{equation}
\epsilon^{\kappa\lambda\mu\nu}\partial^{c(s)}_\lambda \frac{\delta
A_{\min}}{\delta\sigma_{\mu\nu}(c(s))}=(2\kappa-1)\epsilon^{\kappa\lambda\mu\nu}f_\lambda(s)
t_{[\mu}(s)g_{\nu]}(s).
\end{equation}

At this point, $\kappa$ enters as an arbitrary constant, rendering
the Bianchi identity conditional. As becomes apparent, now, from
Eqs. (27), (37) and (40) the arbitrariness of this constant refers
to the fact that we cannot connect uniquely the orthonormal basis
$\{n_M^a(t,\,s)\}$, defined on the surface, with the orthonormal
basis $\{n_\mu^a(s),\,t_\mu(s)\}$ defined on the boundary. It is
important to realize, at the same time that if the
$\vec{g}$-function were known, one could, in principle, compute
its normal variations unambiguously.

In the next section, we explicitly determine the normal variations
of the $\vec{g}$-function for the, non trivial as well as generic,
smooth (Wilson) contour configuration discussed in [13], which
goes by the name of `wavy line' configuration. As we shall see,
the explicit result determines the constant $\kappa$ to be 1/2, as
it bypasses the need of referring to a choice of basis,
$\{n_M^a(t,s)\}$, of the form employed in the analysis in section
3 and leading ro the result expressed by Eq. (67). Given, now,
that $\kappa$, as was introduced in this section, does not depend
on the specific form of the (smooth) Wilson loop boundary, we
consider the relevant result to be an independent way to determine
the value of $\kappa$.

\vspace*{.3cm}

{\bf 5. Wavy line Wilson Contour and the Bianchi Identity}

\vspace*{.2cm}

The wavy line approximation, discussed in [13], is specified by
the assumption that the closed Wilson contours entering the gauge
field-string duality, are described by \be
c_1(s)=s,\,c_i=\phi_i(s),\,\, i=2,\cdot\cdot\cdot, D. \ee with the
transverse components $\phi_i(s)$ being very small. Our objective,
in this section is to expand, to fourth order, $A_{\min}$ in
powers of the $\phi_i$. Following Ref. [13], we begin with the
Hamilton-Jacobi equation for the minimal surface, which, for
$y(s)=y\rightarrow 0$ can be written as \be \frac{\partial
A_{\min}}{\partial y} &=& -{1\over y^2} \int
ds\sqrt{\vec{c}^{'2}-y^4\left(\frac{\delta
A_{\min}}{\delta\vec{c}(s)}\right)^2}\nonumber\\ &=&-{1\over y^2}
\int ds\sqrt{\vec{c'}^{2}-y^4\left(\frac{\delta A_{\min}}
{\delta\vec{\phi}(s)}\right)^2-y^4\left(\vec{\phi}\cdot\frac{\delta
A_{\min}} {\delta\vec{\phi}(s)}\right)^2}, \ee where, for the last
step, we used reparametrization invariance: \be
\vec{c'}\cdot\frac{\delta A_{\min}}{\delta\vec{c}(s)}=0. \ee

To continue we now assume that the minimal area can be cast into
the following general form \be A_{\min}=
\sum\limits_{n=0}^{\infty}{1\over n!}
 \int ds_1\cdot\cdot\cdot ds_n \Gamma_{i_1\cdot\cdot\cdot
i_n} (s_1,\cdot\cdot\cdot,s_n| y) \phi_{i_1}(s_1)\cdot\cdot\cdot
\phi_{i{_n}}(s_n). \ee
 Inserting the above equation into into
(69), expanding the square root and taking the Fourier transform
of both sides one finds \be &&A_{\min}={L_o\over y}+{1\over
2}\int{dp\over 2\pi} \tilde{\Gamma}_2(p|y) \tilde{\phi_i}(p)
\tilde{\phi_i}(-p)\nonumber\\ && +{1\over 8} \int{dp_1\over
2\pi}\cdot\cdot\cdot{dp_4\over 2\pi} \tilde{\Gamma}_4(p_1, p_2,
p_3, p_4|y) \tilde{\phi}_i(p_1)\tilde{\phi}_i(p_2)
\tilde{\phi}_j(p_3)\tilde{\phi}_j(p_4)\nonumber\\ &&\times
2\pi\delta\left(\sum\limits_{i=1}^{4}p_i\right) +{\cal O}(\phi^6)
\ee In the above expression $L_o$ is the length of the contour
(along the direction 1) and we have written
\begin{eqnarray}
&&\quad\quad\Gamma_{i_1 i_2}(s_1,s_2|y)= \delta_{i_1 i_2}
\Gamma_2(p|y) =\delta_{i_1 i_2}  \int{dp\over
2\pi}e^{ip(s_2-s_1)}\tilde{\Gamma}_2(p|y),\nonumber\\&&\quad
 \Gamma_{i_1
i_2i_3i_4}(s_1,s_2,s_3|y)=( \delta_{i_1 i_2} \delta_{i_3 i_4}
+{\rm perms})\Gamma_4(s_2-s_1,s_3-s_1,s_4-s_1|y),\nonumber\\&&
\Gamma_4(s_2-s_1,s_3-s_1,s_4-s_1|y)= \int{dp_1\over 2\pi}
\cdot\cdot\cdot {dp_4\over
2\pi}2\pi\delta\left(\sum\limits_{i=1}^{4}p_i\right)\times\nonumber\\&&
\times e^{i\sum\limits_{i=1}^{4}
p_is_i}\tilde{\Gamma}_4(p_1,p_2,p_3,p_4|y).
\end{eqnarray}

The functions $\tilde{\Gamma}_2$ and $\tilde{\Gamma_4}$ have been
determined in Ref. [13]. Here we present only the leading, finite
part of their expansion in powers of $y$: \be
\tilde{\Gamma}_2=-|p|^3 \ee \be \tilde{\Gamma}_4&=&
\Phi(p_1,p_3)+\Phi(p_1,p_4)+\Phi(p_2,p_3)+\Phi(p_2,p_4)-\Phi(p_1,p_2)-\Phi(p_3,p_4)\nonumber\\
&&-F(p_1,p_2,p_3,p_4|y), \ee with \be
&&F=\left[2\frac{\epsilon_{p_1}\epsilon_{p_2}\epsilon_{p_3}\epsilon_{p_4}+1}{\Delta^3}
+\frac{\epsilon_{p_1}\epsilon_{p_2}\epsilon_{p_3}\epsilon_{p_4}}{\Delta^2}
\left(\sum\limits_{i=1}^{4}{1\over |p_i|}\right)
+\frac{\sum\limits_{i<j}^{}|p_ip_j|}{\Pi\Delta}-
{\Delta\over\Pi}\right]\Pi^2 \nonumber\\ &&\quad\quad
\Phi(p_1,p_2)
=\left[2\frac{\epsilon_{p_1}\epsilon_{p_2}}{\Delta^3}
+\frac{\epsilon_{p_1}\epsilon_{p_2}}{\Delta^2}\left(\frac{1}{|p_1|}+\frac{1}{|p_2|}
\right)+{1\over\Delta} \frac{1}{p_1p_2}\right]\Pi^2 \ee and \be
 \epsilon_p={\rm sign}p,\,\,\Delta= \sum\limits_{i=1}^{4}|p_i|,\,\,\Pi=p_1p_2p_3p_4.
\ee

Given the above relations our first check will refer to the normal
variations of the $\vec{g}$-function. In particular, we shall
prove that no term $\sim \delta'(s_1-s_2)$ appears in the
transverse variation of the $\vec{g}$-function and that the
coefficient of the antisymmetric part is ${1\over 2}$. The
quantity of interest reads \be &&\frac{\delta {g}^a(s_1)}{\delta
\vec{n}^b(s_2)}= n_\mu^a(s_1) n_\nu^b(s_2) \frac{\delta
g_\mu(s_1)}{\delta c_\nu(s_2)}= n_i^a(s_1)
n_j^b(s_2)\times\nonumber\\&&\times \left(\phi_i(s_1)\phi_j(s_2)
\frac{\delta g_i(s_1)}{\delta c_1(s_2)}- \phi_i(s_1) \frac{\delta
g_1(s_1)}{\delta c_j(s_2)}- \phi_j(s_2) \frac{\delta
g_i(s_1)}{\delta c_1(s_2)}+ \frac{\delta g_i(s_1)}{\delta
c_j(s_2)}\right), \ee where we have taken account of the fact that
$c'_\mu n_\mu^a=0\Rightarrow n_1^a=-\phi'_i n_i^a$. It should also
be noted that in the preceeding equation we have written
$s_1=s+{h\over2},\,s_2=s-{h\over 2}$ and for our convenience we
shall eventually integrate both sides over $s$.

Using, now, reparametrization invariance we write \be
g_1=-\phi'_ig_i={1\over \sqrt{\vec{c}^{'2}}}\phi'_i\frac{\delta
A_{\min}}{\delta \phi_i} ,\quad\frac{\delta A_{\min}}{\delta
c_1}=-\phi'_i\frac{\delta A_{\min}}{\delta \phi_i}. \ee
Substituting (79) into (78) and keeping terms up to second order
we find \be \frac{\delta {g}^a(s_1)}{\delta \vec{n}^b(s_2)}&=&
n_i^a(s_1)n_j^b(s_2) \left(\delta'(s_1-s_2)A_{ij}- \frac{\delta^2
A^{(4)}_{\min}}{\delta \phi_i(s_1) \delta
\phi_j(s_2)}\right)+\nonumber\\&& +n_i^a(s_1)
n_j^b(s_2)\Sigma_{ij}+{\cal O}(\phi^4), \ee where \be
A_{ij}=(\phi'_j(s_1) - \phi'_j(s_2)) \frac{\delta
A^{(2)}_{\min}}{\delta\phi_i(s_1)} +\phi'_j(s_2) \frac{\delta
A^{(2)}_{\min}}{\delta \phi'_i(s_2)} -\phi'_i(s_1)\frac{\delta
A^{(2)}_{\min}}{\delta \phi_j(s_1)} \ee and \be
\Sigma_{ij}&=&{1\over 2}\phi'_k(s_1)  \phi'_k(s_1)
 \frac{\delta^2 A^{(2)}_{\min}}{\delta \phi_i(s_1)\delta\phi_j(s_2)}
-\phi'_i(s_1)  \phi'_k(s_1)
 \frac{\delta^2 A^{(2)}_{\min}}{\delta \phi_k(s_1)\delta\phi_j(s_2)}
\nonumber\\&&-\phi'_j(s_2) \phi'_k(s_2)
 \frac{\delta^2 A^{(2)}_{\min}}{\delta \phi_i(s_1)\delta\phi'_k(s_2)}.
\ee In the above equations the expressions $A_{\min}^{(2)}$ and
$A_{\min}^{(4)}$ refer to the minimal area estimation up to second
and fourth order, respectively and can be read from (72). As we
are interested only in the antisymmetric part of the normal
variations (80), we shall ignore the contribution from the term
(82) since it is purely symmetric. It is, now, easy to determine
that \be \frac{\delta^2 A^{(2)}_{\min}}{\delta
\tilde{\phi}_i(k)\delta\tilde{\phi}_j(k')} = \delta_{ij}
2\pi\delta(k+k')\tilde{\Gamma}_2(k) \ee and \be &&\frac{\delta^2
A^{(4)}_{\min}}{\delta \tilde{\phi}_i(k)\delta
\tilde{\phi}_j(k')}= \int{dp_1\over 2\pi}{dp_2\over
2\pi}2\pi\delta(p_1+p_2+k+k') \left(\tilde{M}
(p_1,p_2,k,k')+{1\over 2}\tilde{\Gamma}_4(p_1,p_2,k,k')
\delta_{ij}\right)\times\nonumber\\&&\times \tilde{\phi}_i(p_1)
\tilde{\phi}_j(p_2) +\int{dp_1\over 2\pi}{dp_2\over
2\pi}2\pi\delta(p_1+p_2+k+k')\tilde{\Lambda}(p_1,p_2,k,k')\tilde{\phi}_i(p_1)
\tilde{\phi}_j(p_2) \ee with \be \tilde{M}
\equiv\Phi(p_1,p_2)+\Phi(k,k')-F(p_1,p_2,k,k') \ee and \be
\tilde{\Lambda}\equiv \Phi(k,p_1)+\Phi(k',p_2)-
\Phi(k,p_2)-\Phi(k',p_1). \ee

Taking the Fourier transform of (83) we find \be \frac{\delta^2
A^{(2)}_{\min}}{\delta \phi_i(s_1)\delta\phi_j(s_2)}= \int{dk\over
2\pi}\int{dk'\over 2\pi}e^{-iks_1-ik's_2} \frac{\delta^2
A^{(2)}_{\min}}{\delta \tilde{\phi_i}(k)\delta\tilde{\phi}_j(k')}
=-\delta_{ij}\int{dk\over 2\pi}|k|^3e^{-ik(s_1-s_2)} \ee and
consequently \be \frac{\delta^2
A^{(2)}_{\min}}{\delta\phi_i(s)\delta\phi_j(s_2)} =\int
ds'\Gamma_2(s-s') \phi_i(s'),\,\, \Gamma_2(s)=-\int{dk\over
2\pi}|k|^3e^{-iks}. \ee

One now observes that only the last term on the rhs of (84) gives
an antisymmetric  contribution, so the first one can be ignored.
Employing once again the Fourier transform in (84) one sees that
\be \frac{\delta^2 A^{(4)}_{\min}}{\delta \phi_i\left(s+{h\over
2}\right) \delta \phi_j\left(s-{h\over 2}\right)}&=&\int{dq\over
2\pi}{dk\over 2\pi}
{dp_1\over 2\pi}{dp_2\over 2\pi}2\pi\delta(p_1+p_2+q)\nonumber\\
&&\times e^{-iqs-ihk}\tilde{\Lambda}\left (p_1,p_2,k+{q\over 2},
-k+{q\over 2}\right) \tilde{\phi}_i(p_1) \tilde{\phi}_j(p_2). \ee

Since we are interested in the limit $|h|\rightarrow 0$, we shall
explore the limit $|k|\rightarrow\infty$ in the above relation. As
pointed out already, it is enough for our purposes to examine the
integrated over $s$ version of (78), so we can consider the case
$q=0,\,p_1=-p_2\equiv p$ in the last relation.

Using (76) and (86) we determine \be
\tilde{\Lambda}(p,-p,k,-k)=4\Phi(p,k)&=&
4\left[\frac{\epsilon_p\epsilon_k}{4(|p|+|k|)^3}
+\frac{\epsilon_p\epsilon_k}{4(|p|+|k|)^2} \left({1\over |p|}+
{1\over |k|}\right) +\frac{1}{2(|p|+ |k|) p k}\right]p^4k^4
\nonumber\\
&=&\epsilon_p\epsilon_k|p|^5\left[\frac{x^4}{(1+x)^3}+\frac{3x^3}{1+x}\right],
\ee where, following Ref [13], we have set $x=\frac{|k|}{|p|}$.
Upon taking the limit $x\rightarrow\infty$ we find that \be
\tilde{\Lambda}(p,-p,k,-k)=\epsilon_p\epsilon_k|p|^5\left[3x^2-2x
+{\cal O}\left({1\over x}\right)\right]=3p^3k^2{\rm sign}
k-2p|p|^3k+{\cal O}\left({1\over k}\right). \ee The first term
gives zero contribution in the limit $h\rightarrow 0$, while the
second one leads to \be &&\int ds
\frac{\delta^2A^{(4)}_{\min}}{\delta\phi_i(s+h/2)
\delta\phi_i(s-h/2)}= \int{dk\over 2\pi} \int{dp\over
2\pi}e^{-ihk}\tilde{\Lambda}( p,-p,k,-k) \tilde{\phi}_i(p)
\tilde{\phi}_j(-p)=\nonumber\\&=&-2i\delta'(h) \int{dp\over
2\pi}|p|^3 \tilde{\phi}_i(p)\tilde{\phi}_j(-p)= \delta'(h)\int ds
ds'[ \phi'_i(s) \phi_j(s')- \phi_i(s) \phi'_j(s')]\Gamma_2(s-s')
\nonumber\\ &=& -\delta'(h)\int ds\left[ \phi'_j(s) \frac{\delta
A^{(2)}_{\min}}{\delta\phi_i(s)} -\phi'_i(s) \frac{\delta
A^{(2)}_{\min}}{\delta\phi_j(s)}\right]. \ee

This term exactly cancels the term that appears in (80) in the
limit $h\rightarrow 0$. Thus, it is confirmed, in the framework of
the wavy line approximation, that no term $\propto \delta (h')$
appears in the transverse variation of $\vec{g}$-function. The
first term in (81) reads, in the limit $ h\rightarrow 0$, \be
(\phi'_j(s_1)- \phi'_j(s_2))\frac{\delta A^{(2)}_{\min}}
{\delta\phi_i(s_1)}=h \phi''_j(s))\frac{\delta
A^{(2)}_{\min}}{\delta\phi_i(s)} +{\cal O}(h^2)
=-h\phi''_j(s)g_i(s) +{\cal O}(h^2) +{\cal O}(\phi^4). \ee Thus,
the antisymmetric part of the transverse variation reads \be
-{1\over 2}n_i^a n_j^b(\phi''_ig_j-\phi''_jg_i), \ee which leads
to the conclusion that the value of the constant $\kappa$ that
appears in Eq. (67) of section 5 to be 1/2. As this constant is
independent from the details of the contour which forms the
boundary, we consider the result (94) as valid for an arbitrary
contour and thereby establishes the validity of the Bianchi
identity, equivalently zig zag invariance, for the string- gauge
field connection scenario promoted in Ref [6] by Polyakov.

\vspace*{.3cm}

{\bf 6. Concluding Remarks}

\vspace*{.2cm}

In this work, we have verified an important, from the Physics
standpoint, property of the Wilson loop functional in the
framework of the AdS/CFT -as promoted in Ref. [6] in the
$\lambda\rightarrow\infty$ limit and concretely deliberated in
Refs [13, 14]. In particular, we established a condition for the
validity of the Bianchi identity which, in turn, solidifies the
consistency of the string-gauge field connection in the sense that
it is compatible with the zig zag invariance and hence secures the
validation of Stokes theorem. This very important occurrence has
been explicitly demonstrated in the context of the wavy line
approximation, which sufficiently describes, in a general manner,
a smooth Wilson loop contour. From the Physics point of view, what
we find especially worth noting is that the results in this paper
have been obtained without any knowledge of the
$\vec{g}$-function. The latter is expected to carry all the
dynamics in any particular investigation of interest one wishes to
conduct in the context of the string-based theoretical scheme
adopted in this work. Given, now, that string theory {\it per se}
is formulated in the framework of first quantization, it seems
realistic for one to further pursue the issue of string-gauge
field relation by employing first quantization methodologies on
the field side. The strategy we, specifically, have in mind to
apply for pursuing such a connection would involve, on the gauge
field theoretical side, a first quantization, worldline casting of
gauge field systems, with which we happen to be quite familiar
(see, e.g., Ref. [18] for a typical example). The envisioned focus
of attention in such a study is expected to be placed on the
$\vec{g}$-function in the sense of connecting it with
(non-perturbative) dynamical behaviors in gauge field systems.
Preliminary indications seem to point to a direction according to
which the $\vec{g}$-function is directly linked with the
spin-field interaction dynamics, while perturbative (local)
dynamics are associated the formation of cusps on the Wilson
contour. Such speculations are, of course, subject of concrete
scrutiny, which we intend to explore in the immediate future.




\begin{thebibliography}{99}





\bibitem{Pol 1} A. Polyakov, Phys. Lett. B 82 (1979)247.

\bibitem{PS}J. Polchinski and M. J. Strassler, Phys. Rev. Lett.
88(2002) 031601; JHEP 12 (2003) 0305.

\bibitem{BKD} 22) Z. Bern, L.J. Dixon and D.A. Kosower, Comptes
Rendus Physique 5(2004))55. arXiv: hep-th/0410021

\bibitem{Mald 12}L.F. Alday and J. Maldacena, arXiv:
hep-th/0705.03.03v3

\bibitem{Hooft} G. 'tHooft, Nucl. Phys. B 72(1974) 461.

\bibitem{Polyakov:1975rs} A. Polyakov, Nucl. Phys. B (Proc.
Suppl.) 68(1998) 1.

\bibitem{Mald 1} J. Maldacena, Adv. Theor. Math. Phys. 2 (1998)
231.

\bibitem{Wilson} K. G. Wilson, Phys. Rev. D(1974) 2445.

\bibitem{MM}Yu. M. Makeenko and A. A. Migdal, Phys. Lett. B
97(1980) 253.

\bibitem{Mig-Mak}A. A. Migdal, Phys. Rep. 102 (1983) 199; Yu. M.
Makeenko, {\it Methods of Contemporary Gauge Theory}, Cambridge
Monograph on Mathematical Physics(2002).

\bibitem{Tav} J. N.Tavarez Int. J. Mod. Phys. (1994) 4511.

\bibitem{2HM} M. Hirayama, S. Matsubara, Prog. Theor. Phys
99(1998) 691.

\bibitem{PR 1}A. Polyakov and V. Rychkov, Nucl. Phys. B 581(2000) 116.

\bibitem{PR 2}A. Polyakov and V. Rychkov, Nucl. Phys. B 594(2001) 272.

\bibitem{Pol 3} A. M. Polyakov, Nucl. Phys. B 594(2001) 272.

\bibitem{BGSN}R. A. Brandt, A. Gocksch, M. A. Sato and F. Neri, Phys. Rev. D 26(1982) 3611.

\bibitem{KK}A. I. Karanikas and C. N. Ktorides, JHEP 11(1999) 033.

\bibitem{AKK} S.D. Avramis, A.I. Karanikas, C.N. Ktorides Phys. Rev. D 66 (2002) 045017.


\end{thebibliography}
\end{document}